\documentclass[epj,twocolumn]{webofc}

\usepackage{subcaption}
\usepackage[varg]{txfonts}

\newcommand{\xm}{\relax\ifmmode X_{\mathrm{max}} \else
  $X_{\mathrm{max}}$\fi}
\newcommand{\mxm}{\relax\ifmmode \left<X_{\mathrm{max}}\right> \else
  $\left<X_{\mathrm{max}}\right>$\fi}
\newcommand{\sxm}{\relax\ifmmode \sigma(X_{\mathrm{max}}) \else
  $\sigma(X_{\mathrm{max}})$\fi}
\newcommand{\nm}{\relax\ifmmode N_{\mathrm{max}} \else
  $N_{\mathrm{max}}$\fi}

\begin{document}
\woctitle{ISVHECRI 2018}
\title{Telescope Array Hybrid Composition and Auger-TA Composition Comparison}
\author{
  \firstname{William} \lastname{Hanlon}\inst{1}%
  \fnsep\thanks{\email{whanlon@cosmic.utah.edu}} for the Telescope
    Array Project
}
  
\institute{University of Utah Dept. of Physics and Astronomy \& High
  Energy Astrophysics Institute, 201 James Fletcher Bldg., 115 S 1400
  E, Salt Lake City, UT 84112, USA}

\abstract{Telescope Array (TA) has
  completed analysis of nearly nine years of data measuring the
  atmospheric depth of air shower maximum (\xm) utilizing the TA
  surface detector array and the Black Rock Mesa and Long Ridge
  fluorescence detector stations. By using both the surface array and
  the fluorescence detector, the geometry and arrival time of air
  showers can be measured very precisely providing good resolution in
  determining \xm. \xm{} is directly related to the air shower primary
  particle mass and is therefore important for understanding the
  composition of ultra high energy cosmic rays (UHECRs). UHECR
  composition will help answer questions such as the distance and
  location of their sources. We discuss the experimental apparatus,
  analysis method, and \xm{} data collected. We compare the energy
  dependent distributions of the observed data to detailed Monte Carlo
  simulations of four chemical species, then test which individual
  species are not compatible with the data through an analysis of the
  shapes of the distributions. We also discuss the present state of
  composition analysis and interpretation between the Auger and TA
  experiments. These are the two largest UHECR observatories in the
  world with large exposures and should shed light on UHECR
  composition.}

\maketitle

\section{Introduction}\label{sec:intro}
Ultra high energy cosmic ray (UHECR) mass composition remains a
tantalizing source of inquiry even in the era of very large exposure
detectors. Deeper understanding of the composition of UHECRs will shed
light on which models of propagation and sources are viable. Because
flux in the UHECR regime is low ($J(E) \approx 3 \times 10^{-31}$
eV$^{-1}$~m$^{-2}$~s$^{-1}$~sr$^{-1}$) and is quickly falling with
energy, composition cannot be directly measured on an event-by-event
basis. Large detectors placed on the surface of the Earth, observing
the resultant $N_{2}$ fluorescence of air shower cascades produced by
inelastic collisions of UHECR primaries and air are utilized
instead. Observation of the depth of first interaction could be used
to infer average primary particle mass given a large enough
statistical sample, since this distribution is dependent upon the
cross section with air which is related to particle mass. This method
too is impractical due to the large distances between interaction
depth and detector, low signal-to noise ratio, and relatively small
detector acceptance. Air showers grow in size as a function of
longitudinal depth, which is measured in g/cm$^{2}$, until the average
energy per secondary particle falls below some critical energy. This
depth is called \xm{} and is observable out to several tens of
kilometers by current fluorescence detectors (FDs) in use today
\cite{Carlson:1937zz,Matthews:2005sd}.

Given an ensemble of primaries of a single chemical element with
varying energy, the mean \xm, \mxm, in an energy bin is an increasing
function of the logarithm of primary energy, called the elongation
rate, and it is inversely proportional to the logarithm of the primary
particle mass: $\mxm \propto D \ln(E/A)$, where $D$ is the elongation
rate, $E$ is the energy of the primary particle, and $A$ is the mass
number of the primary. Mixtures of primary elements have $\mxm \propto
\left< \ln A \right>$ \cite{Engel:2011zzb,Kampert:2012mx}. The width
of a distribution of \xm, \sxm, in an energy bin is also related to
the primary particle mass. Intuitively we can guess this result if we
consider a nucleus of mass number $A$ and primary energy $E$, as a
superposition of $A$ nucleons with average energy $E/A$. This is
reasonable in the limit of ultra high energies because the collision
energy is much larger than the binding energy per nucleon. Hadronic
induced showers are more complicated due to effects of elasticity and
multiplicity, so \sxm{} is larger than the naive approximation. For
single elements it is found that \sxm{} of the \xm{}
distribution in an energy bin decreases with increasing primary
mass. For mixtures of elements $\sxm^{2} \propto
D^{2}(\left<\ln^{2}A\right> - \left<\ln A\right>^{2})$, or $\sxm^{2}$
is proportional to the variance of $\ln A$. With these relations among
\mxm{} and \sxm, we can simulate the UHECR flux arriving at a detector and
compare the \xm{} distributions, \mxm, and \sxm{} expected according to
detector exposure to that observed in data. Hadronic air showers are
complicated due to unknown parameters such as cross section,
multiplicity, and elasticity for energies above those that can be
recreated in the controlled experiments such colliders. Therefore
hadronic models play an important role, and are an important
contribution to systematic uncertainty, in understanding UHECR
composition. Model uncertainties will be discussed further in
section~\ref{sec:ta_results}.

\xm{} is not observed by surface detectors (SDs), so, used alone, they
cannot utilize the relationships discussed above to measure primary
mass composition. Work is now progressing in TA and Auger through
seeking to exploit the relationship between primary particle mass and
the muon production in UHECR air showers
\cite{Wahlberg:2009zz,Abbasi:2018wlq}. Muons produced in UHECR induced
air showers are likely to survive to ground level and primaries of
larger mass number produce more muons. This information can
potentially be used to measure the primary mass composition on an
event-by-event basis. One large advantage of this method is that SD
arrays have 100\% duty cycle and run continuously, while FDs run only
on clear, moonless nights providing only a 10\% duty cycle. Good
sensitivity to muons and accurate hadronic models are required to do
this. In particular it is now known that there are large
discrepancies between model predictions of UHECR air shower muon flux
on the ground and that predicted by the most recent hadronic
models. TA and Auger each measure an excess of muons on the ground
versus that predicted by QGSJet~II and EPOS hadronic models
\cite{Aab:2016hkv,Abbasi:2018fkz}. TA has measured $\left<\ln
A\right>$ using nine years of SD data and a multivariate analysis
technique up to $E = 10^{20}$~eV, finding $\left<\ln A\right> = 1.52
\pm 0.08$ \cite{Abbasi:2018wlq}.

This paper will describe the most recent, highest statistics
measurement of \xm{} by Telescope Array in
section~\ref{sec:ta_results}. Section~\ref{sec:ta_auger_comp} will
describe \xm{} measurements of the two largest UHECR observatories and
methods required to compare their results. Section~\ref{sec:summary}
will summarize the details reported in this work.

\section{Telescope Array \xm{} Results}\label{sec:ta_results}
Telescope Array is the largest cosmic ray observatory in the Northern
Hemisphere, covering $\sim 700$~km$^{2}$ in Millard County, Utah
($39.3^\circ$~N, $112.9^\circ$~W, 1400~m asl). Placed within this area
are 507 plastic scintillation surface detectors and three fluorescence
detector stations. Each surface detector is 3~m$^2$ in area and holds
two layers of plastic scintillator 1.25~cm thick. Each layer has 104
wavelength shifting fiber optic cables embedded in them, optically
coupled to a PMT. Each SD contains FADC electronics modules to
digitize signals from the PMTs, GPS modules to record times, and radio
communications equipment to allow communication with one of three
radio towers that monitor the SDs and runs regular diagnostics. When
an SD records a low level trigger, which is simply some signal above
threshold, they communicate this information to the radio towers which
scan multiple low level triggers occurring nearly coincident in
time. When the requirements for an event level trigger are met, the
radio tower broadcasts for a readout of the neighboring SDs and stores
the data for offline analysis \cite{AbuZayyad:2012kk}.

The three FD stations are located to the north
(Middle Drum), southwest (Long Ridge), and southeast (Black Rock) of
the boundary of the SD array, each pointing towards the array
center. The Middle Drum FD station employs 14 FD telescopes repurposed
from the HiRes experiment \cite{AbuZayyad:2000uu}. The Black Rock (BR)
and Long Ridge (LR) FD stations were newly constructed for the TA
experiment. Each of these new FD stations utilize 12 telescopes,
consisting of a 6.8m$^2$ mirror, focusing light onto a photomultiplier
tube (PMT) cluster box containing 256 PMTs arranged in a 16x16 packed
hexagonal array, and data acquisition (DAQ) hardware which includes
FADC readout electronics, pattern recognition modules, and
communications hardware to communicate with a central trigger
distributor which directs readout and storage of data of all telescope
DAQs when the condition for an event trigger is met
\cite{Tameda:2009zza,Tokuno:2012mi}. The twelve mirrors are arranged in a
two ring configuration, with six mirrors in ring 1 observing between
3$^\circ$ - 18$^\circ$ in elevation, and six mirrors in ring 2
observing between 18$^\circ$ - 33$^\circ$. Total azimuthal coverage is
108$^\circ$.

\xm{} data is collected in hybrid mode by searching for time
coincident events in the SD and FD data streams. Hybrid event
observation allows for more accurate reconstruction of the shower
track in the sky because it combines the FD observation of the shower
along with the timing and geometry information of the SD. By using the
SD timing and geometry, the shower angle ($\psi$) in the
shower-detector plane is very well constrained, allowing for a very
accurate measurement of \xm. Monocular reconstruction of showers has
poor \xm{} resolution compared to hybrid and stereo FD reconstruction
methods. Multi-FD reconstruction of events also offers very good \xm{}
resolution, about the same as for hybrid reconstruction, but the
acceptance of stereo events in TA is lower for $E < 10^{19}$~eV due to
the large distance between FD stations. Hybrid reconstruction provides
the largest statistics study of \xm{} in TA.

Previously TA has published results of hybrid \xm{} using the Middle
Drum FD station \cite{Abbasi:2014sfa}. This present paper describes
hybrid \xm{} measurements utilizing the Black Rock and Long Ridge
detectors. Black Rock and Long Ridge are ``twins'' using the same
hardware, same electronics, and are similarly placed close to the
border of the SD array (3~km and 4~km respectively), therefore they
have similar hybrid acceptance. Middle Drum uses smaller mirrors and
is placed about 8~km from the surface array border, giving it a very
different acceptance particularly at low energies. Therefore Middle
Drum reconstruction is not combined with Black Rock and Long Ridge
events for this analysis.

This paper will summarize results of nearly nine years of BR/LR hybrid
\xm{} measurements recently published; further details about the
analysis and results can be found there \cite{Abbasi:2018nun}. The
data examined in this analysis spans the period 27 May 2008 to 29
November 2016, over which 1500 nights of data were collected. Time
matching of BR/LR events and SD events resulted in 17,834 hybrid
candidate events. An event only becomes an accepted hybrid event after
it passes all quality cuts. Cuts on geometry such as track length cut,
shower-detector plane cut, and zenith angle cut are used to ensure
accurate reconstruction of the shower track. Cuts on fluorescence
profile $\chi^2$, \xm{} bracketing, and good weather nights ensure the
shower profile is well observed and \xm{} is well measured. An energy
cut is also applied to ensure that no events with reconstructed energy
below $10^{18.2}$~eV are accepted. This cut is imposed because of
rapidly falling acceptance near this energy. After all quality cuts
are applied 3330 events remain as accepted hybrid events for $E \geq
10^{18.2}$~eV. To ensure the validity of our analysis TA makes
extensive use of a detailed Monte Carlo simulation which includes
hourly and nightly databases of important runtime parameters such as
PMT pedestals, PMT gains, mirror reflectivity, and atmospheric
profile. We simulate the running conditions of each night's data that
is collected to provide a simulated data sample approximately 10x that
collected in data. Data and Monte Carlo are packed into the same data
format and reconstructed using the same programs. Because of model
dependencies and primary particle dependencies on observables, Monte
Carlo is thrown for different models and primaries. In this work we
compare TA \xm{} data to QGSJet~II-04 \cite{Ostapchenko:2010vb}
protons, helium, nitrogen, and iron. We then perform data/Monte Carlo
comparisons on observables that are important to good hybrid
reconstruction of \xm. Reconstruction bias and resolution is also
checked by examining the difference between thrown and reconstructed
values of important reconstruction parameters. \xm{} reconstruction
bias is -1.1, -3.3, -3.8, and -3.8~g/cm$^2$ for QGSJet~II-04 protons,
helium, nitrogen, and iron respectively for $E \geq
10^{18.2}$~eV. \xm{} resolution is 17.2~g/cm$^2$ for protons and
13.2~g/cm$^2$ for iron. Energy bias ranges from 1.7\% for protons to
-6.5\% for iron, with energy resolution no worse than 6\% for any of
the four species examined. Resolution and bias of important geometric
observables are very good as expected for hybrid
reconstruction; angular resolutions are $< 1^\circ$ and resolutions on
$R_{\textrm{p}}$ (shower impact parameter) and core location are $\sim
50$~m.

We can compare the \xm{} distributions of the data with that predicted
by the Monte Carlo to see what elements best describe our
observations. It is not clear what fraction of UHECRs in an energy bin
are mixtures of different elements. For this work TA only compares
data to the four individual QGSJet~II-04 elements we simulated. This
can be done by comparing the first two moments of the observed \xm{}
distributions in an energy bin, e.g., \mxm{} and \sxm. Simplifying
entire distributions to their first and second moments is problematic
in the case of \xm{} distributions though. \xm{} distributions are not
normal distributions and may exhibit significant skew depending upon
the primary element being examined. Light elements such as protons and
helium have large positive skew (right tail is longer), while heavy
elements such as iron have a much less pronounced tail. This in and of
itself may not be a problem when sufficient statistics are available
in data. UHECR flux falls rapidly with energy though and appears to be
cutting off above $10^{19.7}$~eV \cite{AbuZayyad:2012ru}. This
requires that UHECR observatories have very large exposures to provide
sufficient statistics to generate a distribution of events that is not
susceptible to statistical biases when measuring their means and
widths. Hybrid reconstruction of \xm{} is limited in its acceptance by
requiring events fall within the boundaries of its SD array and TA's
SD reconstruction is limited to zenith angles of 60$^\circ$. These two
constraints reduce the total acceptance of hybrid reconstructed events
as energy increases because \xm{} occurs closer to, or even in, the
ground. Hybrid reconstruction of \xm{} requires that \xm{} be within
the field of view of the FD (\xm{} bracketing). As energy increases
only tracks with increasing zenith angle have the potential to have
sufficient track length to reach shower maximum in the
atmosphere. Smaller zenith angle acceptance though reduces the FD
field of view for showers that also arrive relatively close by
due to the constraint imposed by the simultaneous SD detection as
well. Standard FD-only observation has an aperture that grows with
energy beyond the boundaries of the SD array, whereas hybrid can only
grow up until the acceptance of the far side of the SD is fully
efficient. TA SD reconstruction is currently limited up to zenith
angles of 60$^\circ$ due to the difficulty of reconstructing the
footprint of the shower at such large angles. Stereo FD reconstruction
can address some of these problems. Stereo FD has equally good \xm{}
resolution as hybrid, is not limited to a smaller field of view at the
highest energies, and has better acceptance than hybrid reconstruction
at the highest energies. Because of the potential loss of the most
deeply penetrating events at the highest energies, the potential for
bias in \mxm{} and \sxm{} must be considered. We present the results
of the first two moments of the observed \xm{} distributions, but we
also employ a morphological test of the shapes of the \xm{}
distributions and calculate the probability that a given element is
compatible with our observations at the 95\% confidence level.

\begin{figure}
  \centering
  \includegraphics[clip,width=0.95\linewidth]{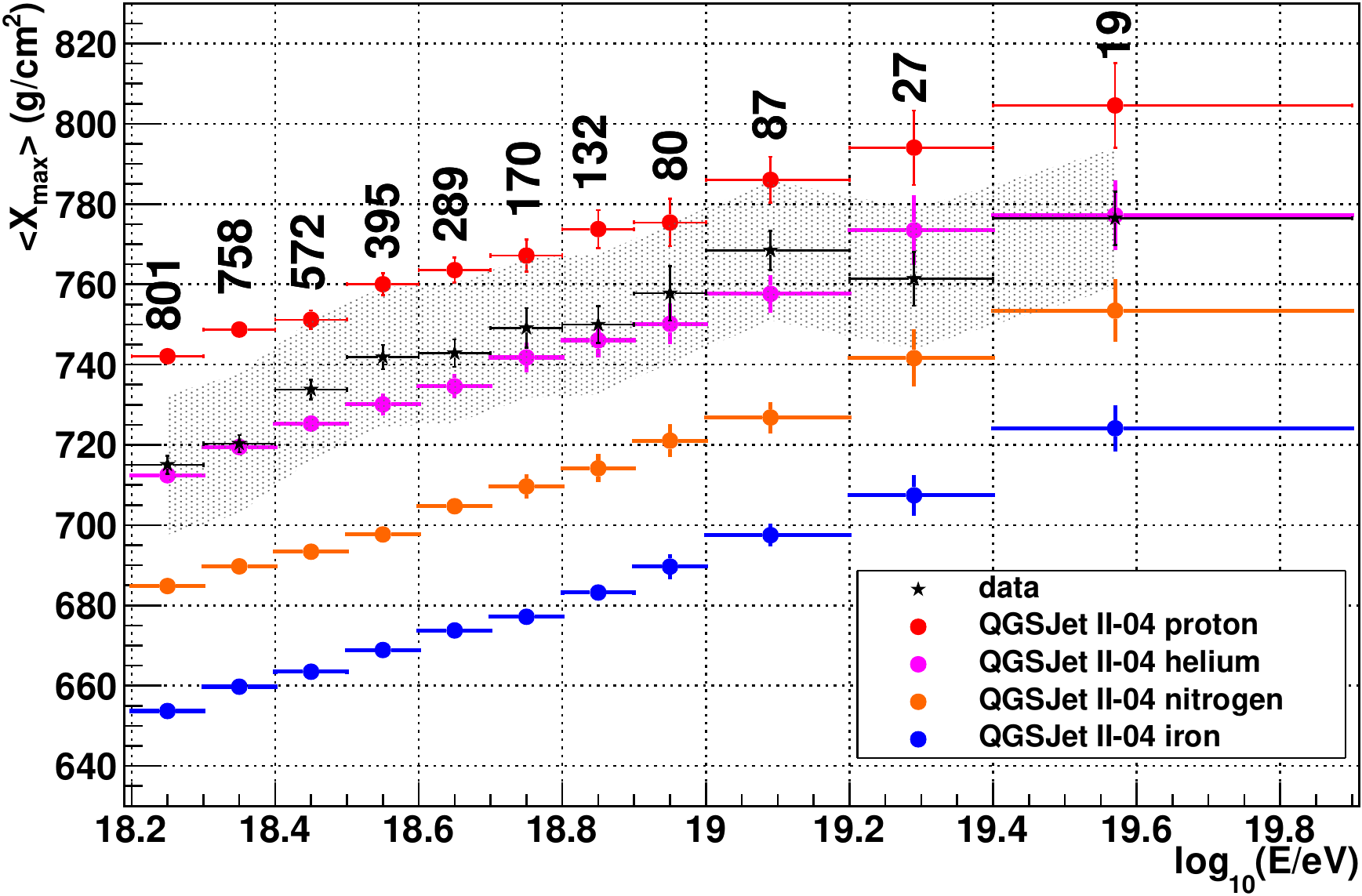}
  \caption{Mean \xm{} as a function of energy as observed by Telescope
    Array in BR/LR hybrid mode over nearly nine years of data
    collection. The numbers above the data points indicate the number
    of events observed. The gray band is the systematic uncertainty of
    this analysis. Reconstructed Monte Carlo of four different primary
    species generated using the QGSJet~II-04 hadronic model are shown
    for comparison.}
  \label{fig:data_xmax}
\end{figure}

Figure~\ref{fig:data_xmax} and table~\ref{tab:data_xmax} summarize the
observed \xm{} for nearly nine years of hybrid operations utilizing
the BR and LR FD detectors in coincidence with the SD
array. Figure~\ref{fig:data_xmax} also shows the predicted \mxm{} for
the four QGSJet~II-04 species simulated as well. \mxm{} of the data
differs greatly from nitrogen and iron, more closely resembling \mxm{}
of protons or helium. Notice that as energy increases the size of the
energy bins must be increased to attempt to capture sufficient events
to make a good measurement of \mxm{} and \sxm. Having only 27 or 19
events in a bin though, such as we have for $E \geq 10^{19.2}$~eV,
means we are susceptible to bias.

\begin{table}
  \centering
  \begin{tabular}{rrrrrr}
    $E_{\mathrm{low}}$ &$\left<E\right>$ &$E_{\mathrm{high}}$ &$N_{\mathrm{data}}$ &\mxm &\sxm\\
    \hline
    18.20 &18.25 &18.30 &801 &$715 \pm 2^{+17}_{-17}$ &$63 \pm 2^{+3}_{-4}$\\
    18.30 &18.35 &18.40 &758 &$720 \pm 2^{+17}_{-17}$ &$59 \pm 2^{+4}_{-4}$\\
    18.40 &18.45 &18.50 &572 &$734 \pm 2^{+17}_{-17}$ &$58 \pm 2^{+4}_{-4}$\\
    18.50 &18.55 &18.60 &395 &$742 \pm 3^{+17}_{-17}$ &$61 \pm 3^{+4}_{-4}$\\
    18.60 &18.65 &18.70 &289 &$743 \pm 3^{+17}_{-17}$ &$58 \pm 3^{+4}_{-4}$\\
    18.70 &18.75 &18.80 &170 &$749 \pm 5^{+17}_{-17}$ &$65 \pm 6^{+3}_{-4}$\\
    18.80 &18.85 &18.90 &132 &$750 \pm 5^{+17}_{-17}$ &$52 \pm 5^{+4}_{-4}$\\
    18.90 &18.95 &19.00 &80  &$758 \pm 7^{+17}_{-17}$ &$61 \pm 8^{+4}_{-4}$\\
    19.00 &19.09 &19.20 &87  &$769 \pm 5^{+17}_{-17}$ &$46 \pm 4^{+5}_{-5}$\\
    19.20 &19.29 &19.40 &27  &$761 \pm 7^{+17}_{-17}$ &$35 \pm 4^{+6}_{-7}$\\
    19.40 &19.57 &19.90 &19  &$777 \pm 7^{+17}_{-17}$ &$29 \pm 4^{+7}_{-9}$ \\
  \end{tabular}
  \caption{\mxm{} and \sxm{} observed for nearly 9 years of data by
    Telescope Array in BR/LR hybrid collection mode. Energy is in
    units of $\log_{10}(E/\mathrm{eV})$ and \mxm{} and \sxm{} are in
    g/cm$^{2}$.}
  \label{tab:data_xmax}
\end{table}

\begin{figure}
  \centering
  \includegraphics[clip,width=0.98\linewidth]{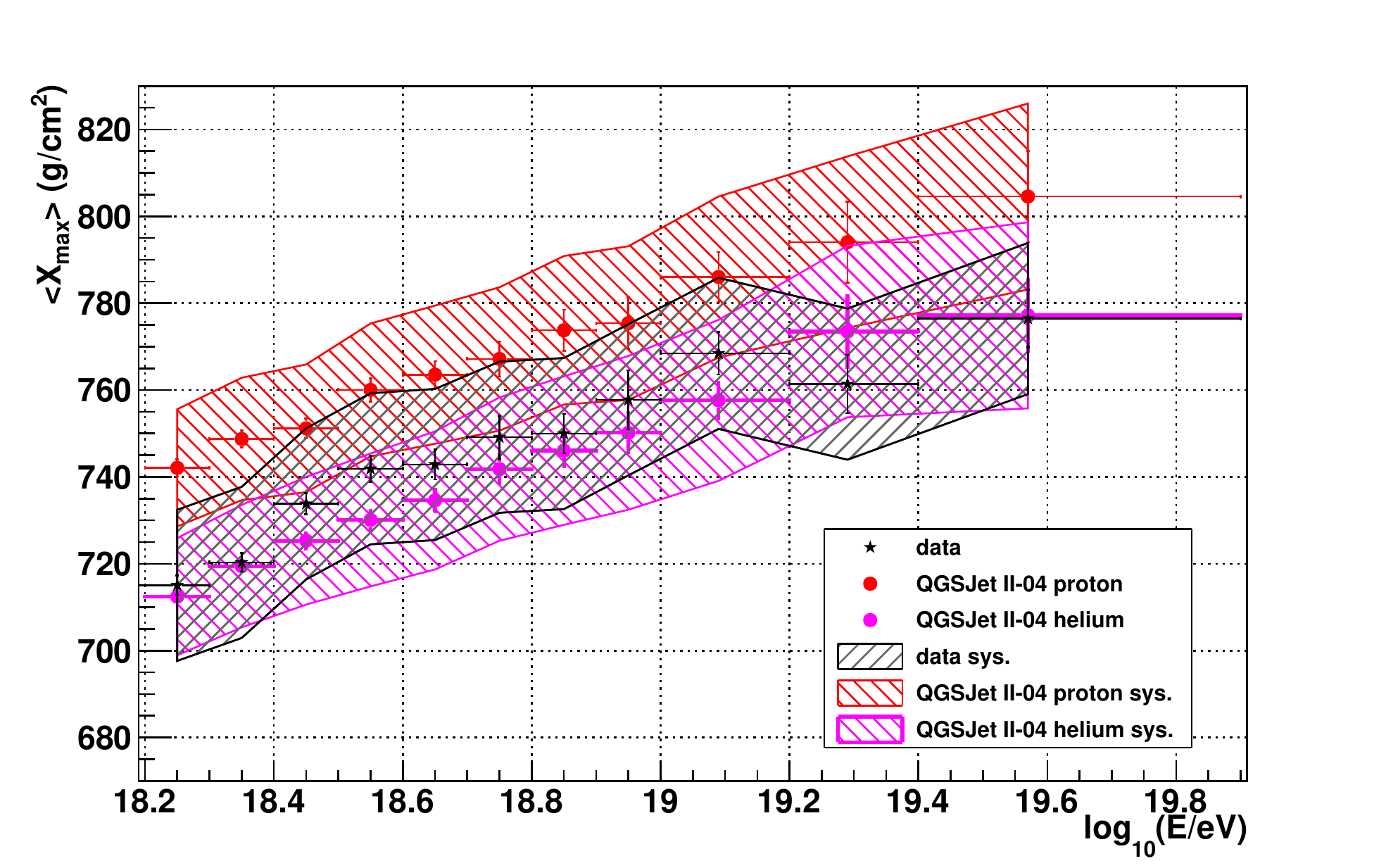}
  \caption{TA BR/LR observed hybrid \mxm{} and predictions for
    QGSJet~II-04 protons and helium. The black band around the data is
  calculated systematic uncertainty of the data. The colored bands
  around the proton and helium predictions are estimated uncertainties
  in the QGSJet II-04 model.}
  \label{fig:xmax_model_uncertainty}
\end{figure}

We can measure the compatibility of the data with Monte Carlo by
performing a maximum likelihood fit between their \xm{}
distributions. To do this we need to address the potential of
systematic uncertainties in our reconstruction and potentially in the
models we are comparing against. We do this introducing a parameter
$\lambda$ by which we shift the data \xm{} distribution and finding
the $\lambda$ which gives the best likelihood between the data and
Monte Carlo for a given energy bin and chemical element. The
uncertainties in \mxm{} in hadronic models have been estimated to
range from $\sigma(\mxm) = \pm 3$~g/cm$^2$ at $10^{17.0}$~eV to
$\sigma(\mxm) = \pm 18$~g/cm$^2$ at $10^{19.5}$~eV
\cite{Abbasi:2016sfu}.  Figure~\ref{fig:xmax_model_uncertainty} shows
the estimated systematic uncertainty in QGSJet~II-04 protons and
helium relative to TA \mxm{} and there is significant overlap between
them. These uncertainties arise mainly from the relatively large
unknown contributions from cross section, multiplicity, and elasticity
in hadronic models that must be extrapolated from current collider
energies up to the energies of UHECRs \cite{Ulrich:2010rg}.

\begin{figure}
  \centering
  \includegraphics[clip,width=0.95\linewidth]{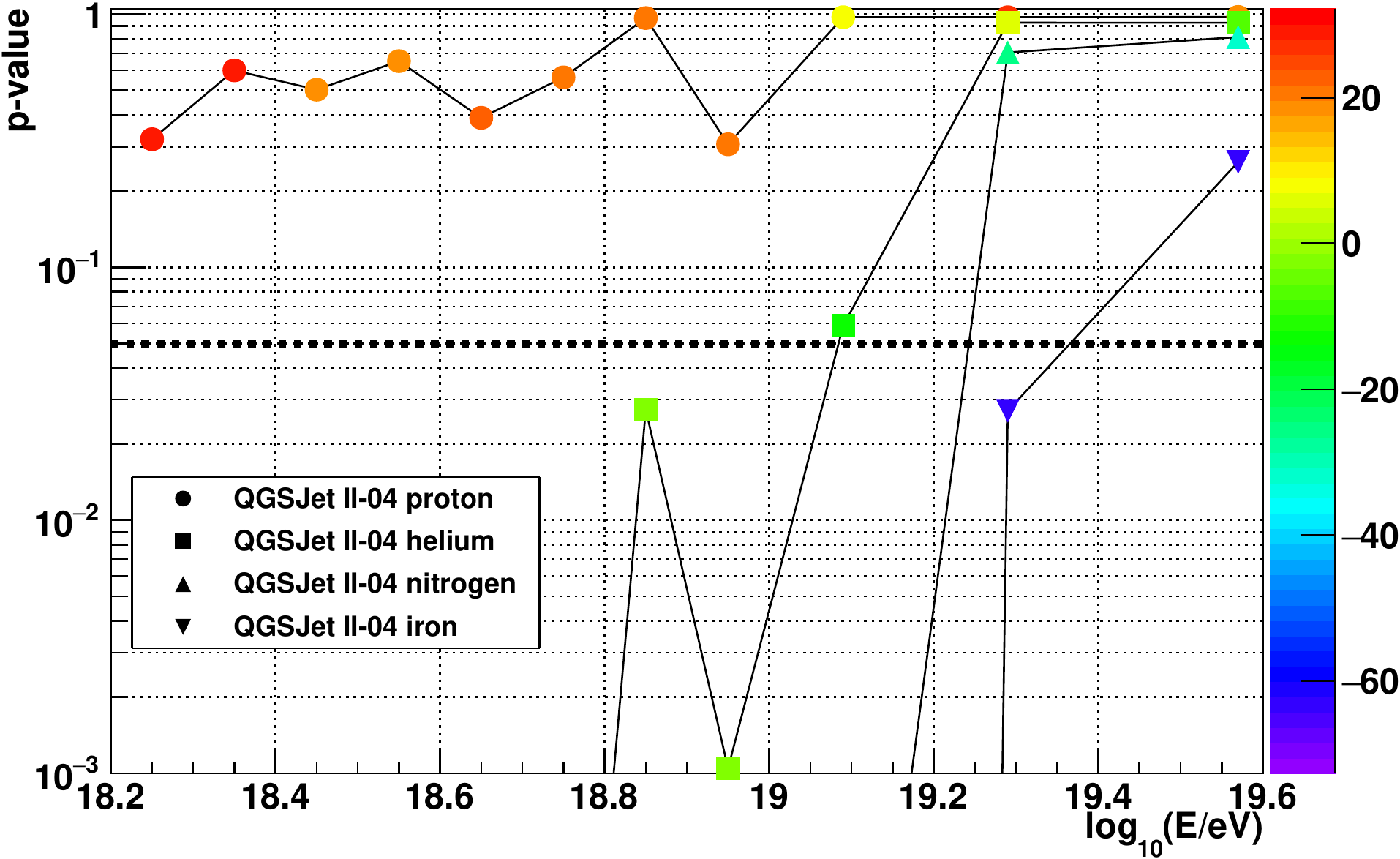}
  \caption{Unbinned maximum likelihood test on observed and simulated
    QGSJet~II-04 \xm{} distributions after systematic shifting of the
    data to find the best log likelihood. Each point represents the
    probability of measuring a log likelihood more extreme than that
    observed in the data after it is shifted by the best $\Delta
    X_{\mathrm{max}}$. The color of the point indicates the $\Delta
    X_{\mathrm{max}}$ measured in g/cm$^{2}$ required to find the
    maximum log likelihood value. The dashed line at $p$-value = 0.05
    indicates the threshold below which the data is deemed
    incompatible with the Monte Carlo at the 95\% confidence level.}
    \label{fig:dataMCMLTest}
\end{figure}

We then calculate the chance probability of observing a likelihood at
least as extreme as we find using the data. Using a critical value of
5\%, we say that if the probability of observing a likelihood measured
for the data and a given species is less than 5\%, then the data is
not compatible with the model at the 95\% confidence level. If, on the
other hand, the probability is greater than 5\% we say that we cannot
rule out that species as being the same as that which we have
observed. We note that this is a simple test against single chemical
elements, and does not test for compatibility for mixtures of
elements. Our results do not preclude the possibility for mixtures of
elements given that they occur in the correct ratios that would allow
their distributions to mimic those we observe in our
data. Figure~\ref{fig:dataMCMLTest} shows the results of these tests
for QGSJet~II-04 protons, helium, nitrogen, and iron. For all energies
protons are found compatible with TA data. Helium does not become
compatible with our data until $E > 10^{19.2}$~eV. Nitrogen becomes
compatible for $E > 10^{19.2}$~eV and iron is compatible for $E >
10^{19.4}$. It may seem problematic that our \xm{} data is
simultaneously compatible with \xm{} distributions whose shapes vary
as widely as protons and iron at the highest energies. This reflects
the lack of sufficient statistics to accurately measure the shapes of the
distributions. In particular the tails of the distributions of the
light elements disappear in these energy bins potentially because the
\xm{} distribution is shifted deeper as it is approaching the limit of
TA's reconstruction acceptance for such deeply penetrating events,
even for simulated data. For iron though,
figure~\ref{fig:dataMCMLTest} shows that the shift in \xm{} required
to find the maximum likelihood requires a very large shift,
approximately 60~g/cm$^2$, making it unlikely that iron is a truly
reasonable match between the data and the Monte Carlo.

\section{Comparison of Auger and Telescope Array
  \xm}\label{sec:ta_auger_comp} In the Southern Hemisphere, the Pierre
Auger Observatory is the largest cosmic ray observatory currently in
operation. It is located near Malarg\"{u}e, Argentina. It uses 1660
water Cherenkov detectors covering 3000~km$^2$ and 24 fluorescence
telescopes at four FD stations overlooking the SD array
\cite{ThePierreAuger:2015rma}. Auger also uses hybrid observation to
measure \xm.

Comparison of \xm{} results between Auger and TA must account for the
way each group analyzes their data. Auger reconstructs \xm{}
distributions as ``seen in the atmosphere'', meaning they select
events such that their final \xm{} distributions for an energy bin are
not biased by acceptance. This is done primarily through cuts on the
field of view of the FD telescopes. For each energy bin, the upper and
lower bound of the acceptable field of view is found by finding that
range that does not allow true \xm{} to deviate by more than
5~g/cm$^2$. All observed showers must fall within this range
otherwise it is not accepted. Other cuts are applied as well to ensure
all events are high quality events. The effect of such a procedure is
that the observed \xm{} distributions will look like a distribution of
simulated \xm{} before it is distorted through acceptance of the detector
\cite{Aab:2014kda}. Model dependence is still an issue and data
must be compared to thrown models if one wishes to understand the
distribution of chemical elements in the primary spectrum. Telescope
Array reconstructs \xm{} distributions as ``seen by the detector''. TA
imposes minimal cuts in an attempt to collect as many events as
possible. In this approach, biases are incurred into a thrown \xm{}
distribution, caused by loss of events, e.g., high energy small zenith
angle events that achieve shower maximum outside the field of view of
the detector. This can have the effect of truncating tails in the
reconstructed \xm{} distributions and shifting their means and
widths. We can still compare our data to simulations because we
utilize a very detailed Monte Carlo simulation that informs us of our
biases and distorts thrown distributions accordingly.

\begin{figure}
  \centering
  \includegraphics[clip,width=0.98\linewidth]{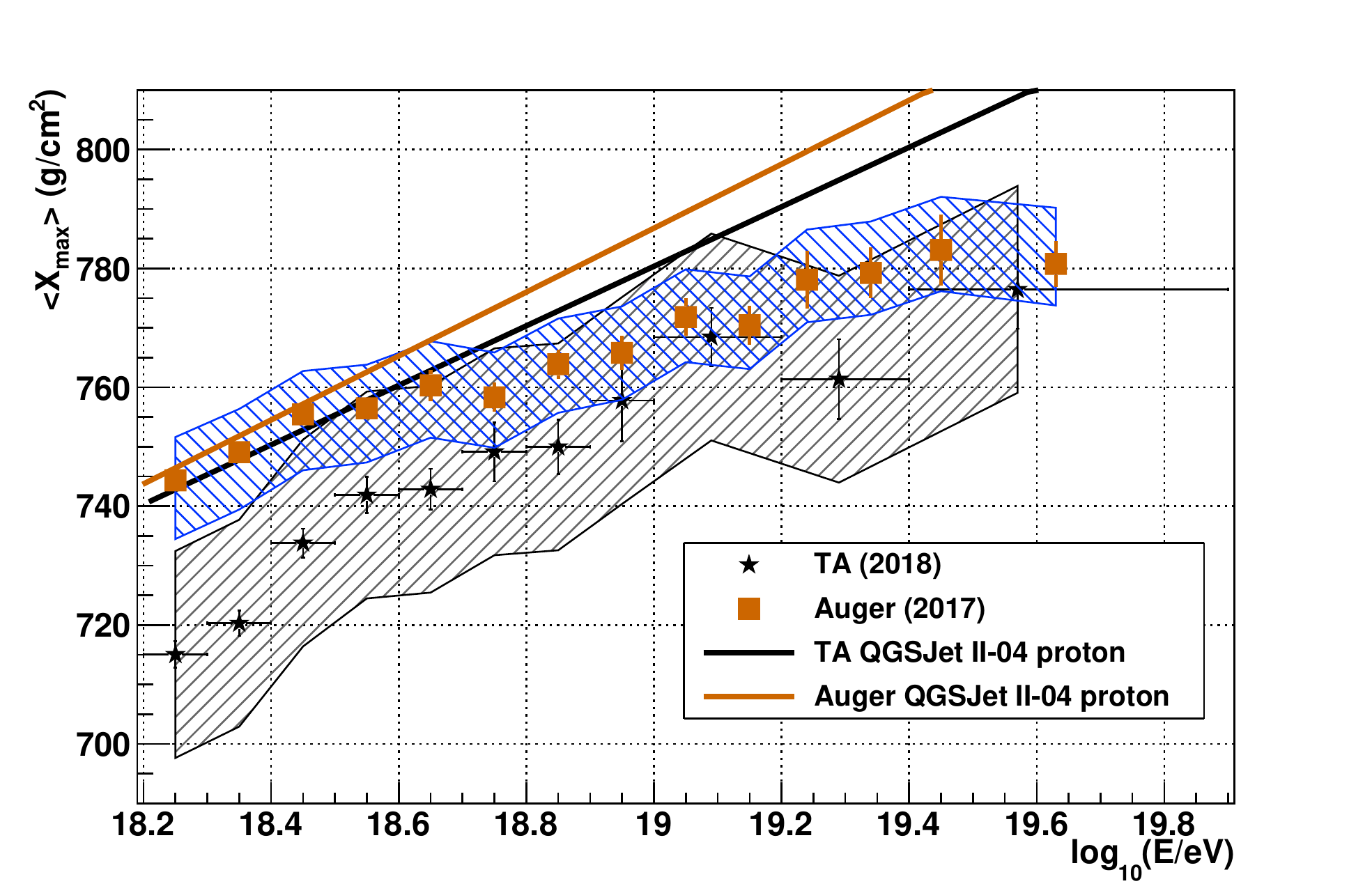}
  \caption{Most recent \mxm{} measurements from TA (2018)
    \cite{Abbasi:2018nun} and Auger (2017)
    \cite{Bellido:2017cgf}. Systematic uncertainties for each
    measurement are indicated by the bands around the data points. The
    orange solid line and black solid line show each experiment's
    prediction to compare against their data. Note that Auger's proton
    prediction is unbiased by acceptance and reconstruction effects,
    while TA's is biased.}
  \label{fig:ta_auger_mean_xmax_data}
\end{figure}

Even given this difference in approach to \xm{} reconstruction, the
data of TA and Auger do not look particularly different given each
experiment's uncertainties. Figure~\ref{fig:ta_auger_mean_xmax_data}
shows the latest \xm{} data collected by TA and Auger. Using nearly
twelve years of data Auger has collected 10,570 events above $E \geq
10^{18.2}$~eV compared to TA's 3330 events
\cite{Bellido:2017cgf}.  Auger's
\mxm{} is unbiased by the detector and TA's \mxm{} is biased. Without
addressing this difference \mxm{} cannot be compared in such a simple
matter. Figure~\ref{fig:ta_auger_mean_xmax_data} also shows as
reference the prediction each experiment makes for QGSJet~II-04
protons. Auger's \mxm{} is consistent with their unbiased prediction
with protons up to $10^{18.7}$~eV, after which the data appears to
signal an increase in primary mass. TA's \mxm{} is consistent with
their prediction with protons within systematic uncertainties for $E
\geq 10^{18.5}$~eV. These results are often interpreted as Auger sees
an evolution in composition to heavier primaries with energy, while TA
sees composition compatible with light primaries such as
protons. 

\begin{figure}
  \centering
  \includegraphics[clip,width=0.98\linewidth]{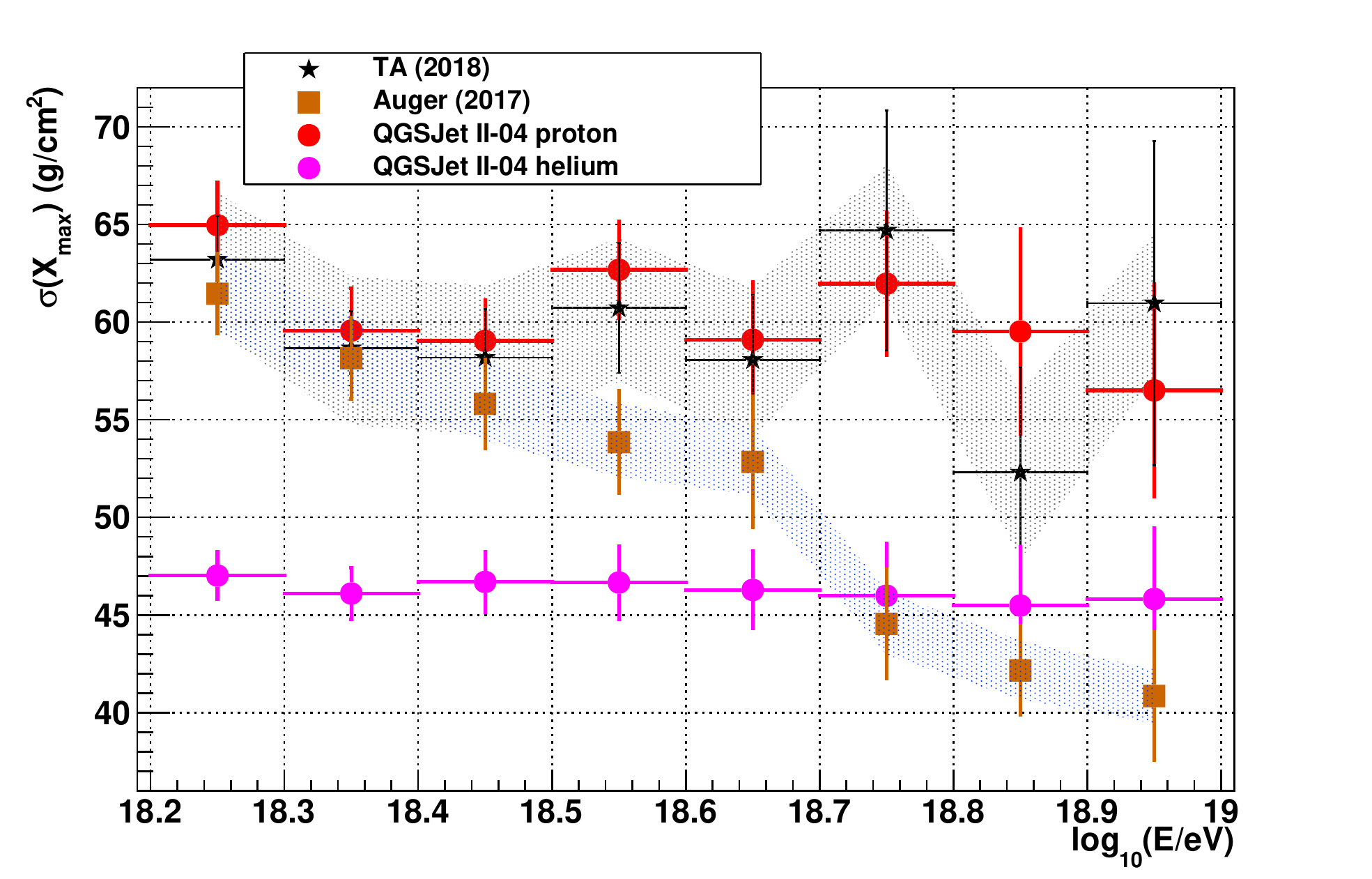}
  \caption{Most recent \sxm{} measurements from TA (2018)
    \cite{Abbasi:2018nun} and Auger (2017)
    \cite{Bellido:2017cgf}. Systematic uncertainties for each
    measurement are indicated by the bands around the data points. The
    TA prediction for \sxm{} of QGSJet~II-04 protons and helium are
    also shown. Auger observes a narrowing of \sxm{} for $E \geq
    10^{18.5}$~eV, while TA does not. Narrowing of \sxm{} is signature
    of composition increasing in mass.}
  \label{fig:ta_auger_sigma_xmax_data}
\end{figure}

\sxm{} for each experiment also shows results consistent with
these interpretations. Figure~\ref{fig:ta_auger_sigma_xmax_data} shows
recent measurements of \sxm{} for both experiments. Below
$10^{18.5}$~eV Auger and TA observe the same \sxm{}. Above this
energy, Auger sees a narrowing of the \xm{} distributions, which may be a
sign of increasing primary particle mass.

Claims of disagreement between TA and Auger composition usually arise
from naively interpreting the data as displayed in
figures~\ref{fig:ta_auger_mean_xmax_data} and
\ref{fig:ta_auger_sigma_xmax_data}, without regard to the differences
in TA and Auger reconstruction. To address the potential confusion
arising from how to properly compare TA biased and Auger unbiased
\xm{} distributions, a Composition Working Group was formed in 2012 to
allow \xm{} analysis experts in each experiment to work together to
determine the best way to make comparisons of each experiment's data
\cite{Barcikowski:2013nfa}. In the years following, a procedure was
developed to allow the direct comparison of Auger and TA data. The
procedure is as follows: Auger fits their data to an ad-hoc mixture of
proton, helium, nitrogen, and iron, computing the fractions of each as
a function of energy, Auger provides these fractions to TA, TA
performs a Monte Carlo simulation of the mixture, then compares the
reconstructed mix to TA data. Because Auger's data is unbiased, a fit
to their data is a fit to distributions generated by simulation
without detector acceptance distorting them. By reconstructing that
unbiased mixture through the TA analysis chain, and exposing the
distributions to TA acceptance and reconstruction bias, this distorts
the thrown distributions in a way that is consistent with how true
distributions in nature are distorted.

\begin{figure}
  \centering
  \includegraphics[clip,width=0.98\linewidth]{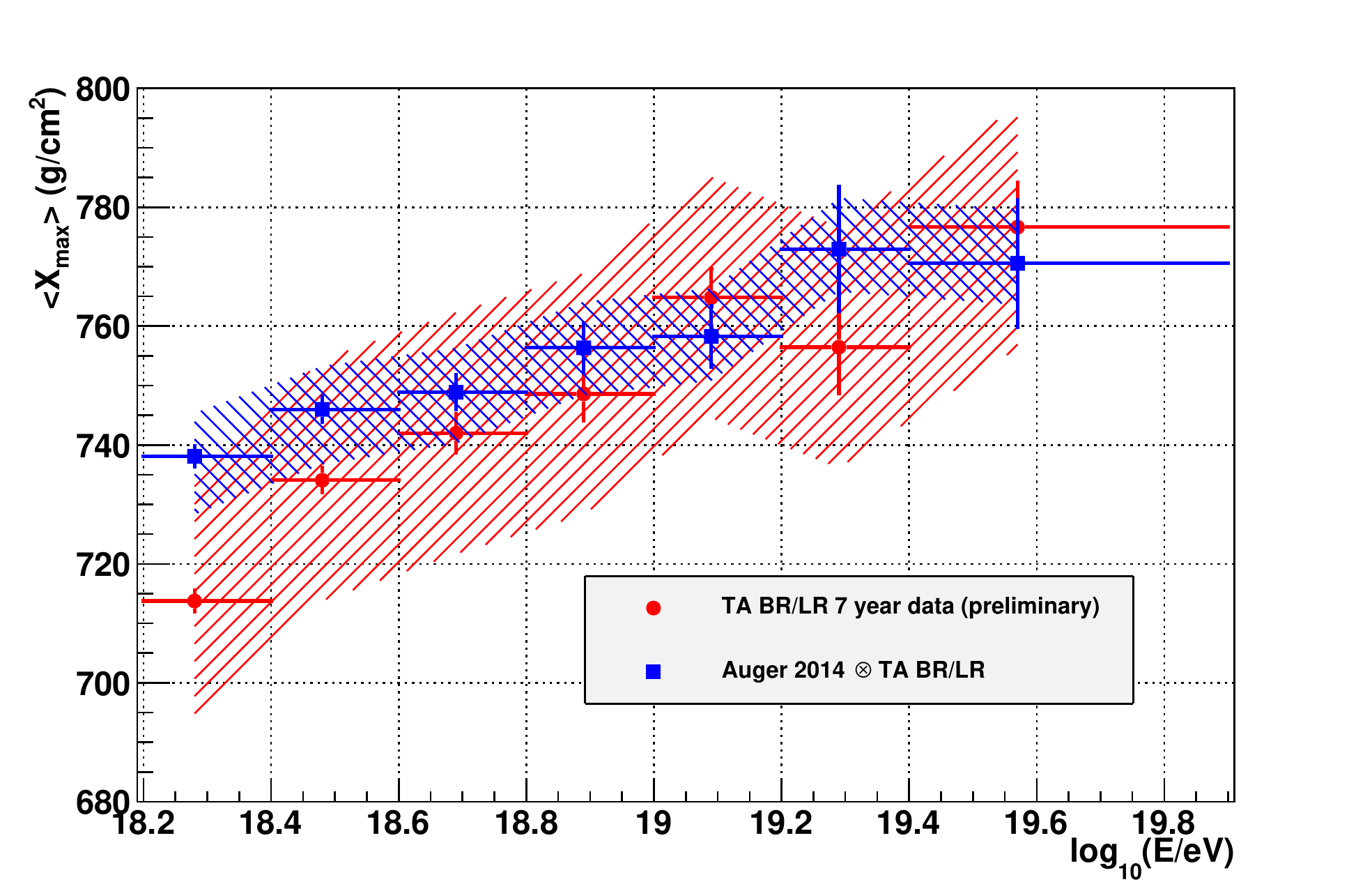}
  \caption{Comparison of Auger and TA \xm{} measurements. TA data is
    that observed by the detector in the field. The Auger 2014
    $\otimes$ TA BR/LR mixture is constructed by fitting Auger data to
    a mixture of four primary elements, throwing that energy dependent
    mixture in Monte Carlo, then reconstructing the mixture in the
    same way TA data is reconstructed. Within systematic uncertainties
  Auger and TA \mxm{} are in agreement.}
  \label{fig:uhecr_2016_mix}
\end{figure}

The issue of which hadronic model best agrees with data must be
addressed as well. Auger has fit its data to a few different models
and finds best agreement with EPOS-LHC
\cite{Aab:2014aea,Pierog:2013ria}. In 2016, Auger and TA presented
results of seven years of TA BR/LR hybrid \xm{} data and a simulated
mixture fit to Auger data and reconstructed through TA's analysis
routines \cite{Hanlon:2018dhzx}. Figure~\ref{fig:uhecr_2016_mix} shows
the result of employing the procedure previously
described. Reconstruction of the mix that fits Auger data effectively
biases it in the way the TA detector would see it. The figure shows
that within systematic uncertainties Auger and TA are in good
agreement in the measurement of \mxm{}.

This analysis was extended further in 2017 \cite{deSouza:2017wgx}. In
that study, the Auger mix was simulated by TA and reconstructed
through TA analysis software, then the reconstructed \xm{}
distributions of the mixture were tested against the TA \xm{} data
distributions using both the Kolmogorov-Smirnov (KS) test and the
Anderson-Darling (AD) test. This procedure allowed us to test for
compatibility of entire \xm{} distributions, not just \mxm{} and
\sxm{}, which is a more powerful and empirical measure. In each case
the tests allow one to calculate the probability that both
distributions were sampled from the same parent distribution. The KS
test is most sensitive near the peak of distributions under test,
while the AD test is more sensitive to the tails of the distributions
\cite{Porter:2008mc}. Sensitivity to the tails is important for skewed
distributions such as \xm. To test the compatibility of the
distributions, in an energy bin the Auger mix \xm{} distribution
reconstructed through the TA analysis chain was sampled according to
TA data statistics, and the KS and AD test statistics were computed for
this sample and the \xm{} distribution of the mixture, called
P1$^{\textrm{MC}}$. This was done 100,000 times and provides a
distribution of probabilities for compatibility of the Auger mix in
that energy bin. The probability that TA data and the Auger mix were
sampled from the same distribution was also calculated and called
P1$^{\textrm{data}}$. The compatibility probability of TA data and the
Auger mix, P2, was then found by computing the probability to find
P1$^{\textrm{MC}}$ larger than P1$^{\textrm{data}}$. Because the Auger
mix is generated using the Auger data, this provides a path for a
direct measure of compatibility of Auger and TA \xm.

\begin{figure}
  \centering
  \includegraphics[clip,width=0.98\linewidth]{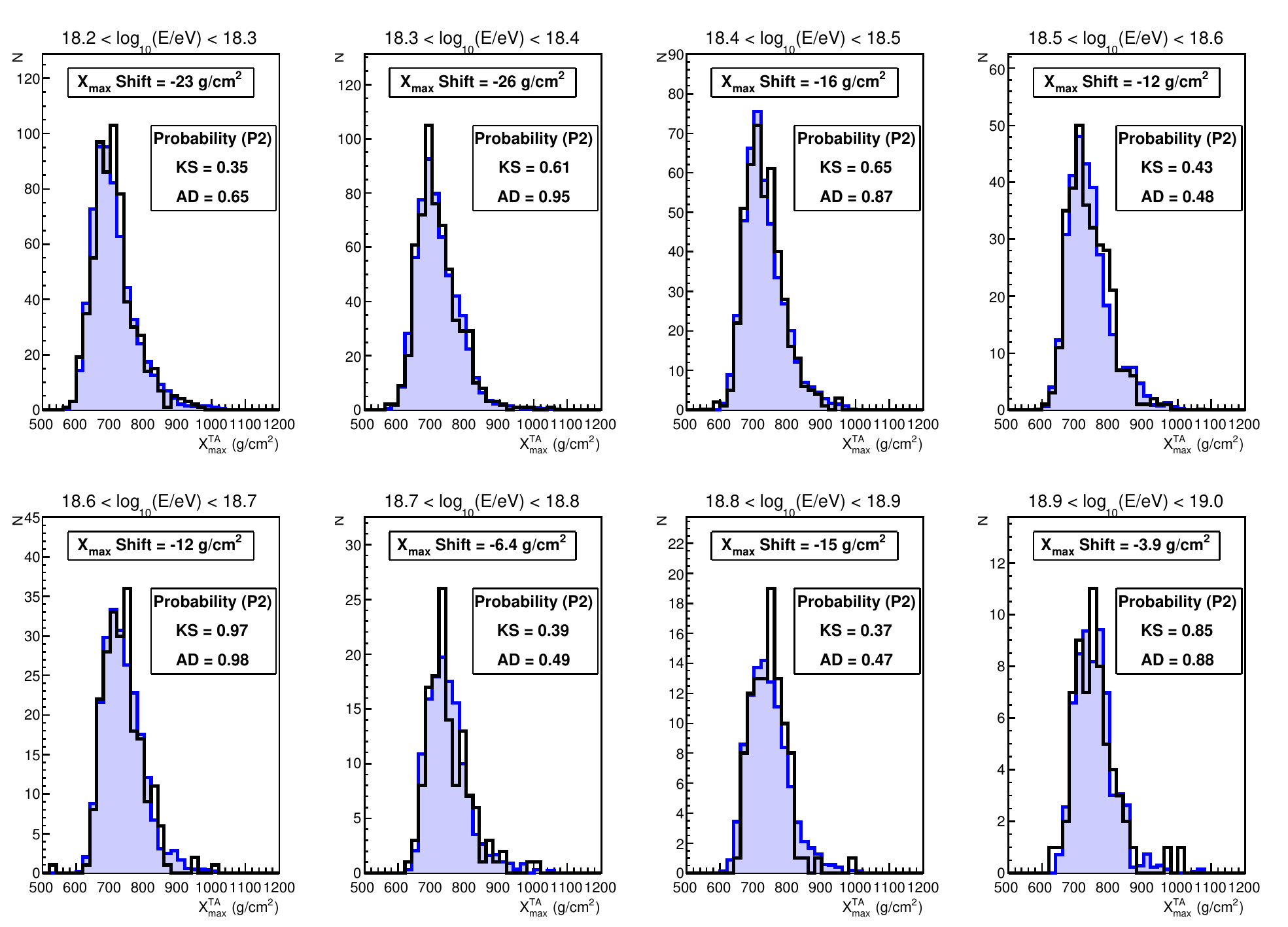}
  \caption{TA data \xm{} distributions (black line) and Auger mix
    \xm{} distributions (solid blue) used to test compatibility of TA
    and Auger data. TA data was allowed to be systematically shifted
    so that the means of the \xm{} distributions matched. The KS and
    AD tests measure the probability of compatibility based upon the
    shapes of the distributions.}
  \label{fig:auger_ta_dist}
\end{figure}

Without systematic shifting of the TA data, only marginal
compatibility was found between Auger and TA data. When TA \xm{}
distributions were allowed to vary systematically in an energy bin so
that the means of the distributions matched,
between $~ 20 - 30$~g/cm$^2$, the compatibility probabilities
indicated agreement for both two-sample
tests. Figure~\ref{fig:auger_ta_dist} shows how the shapes of the
TA and Auger mix distributions agree after shifting.

\begin{figure}
  \centering
  \includegraphics[clip,width=0.98\linewidth]{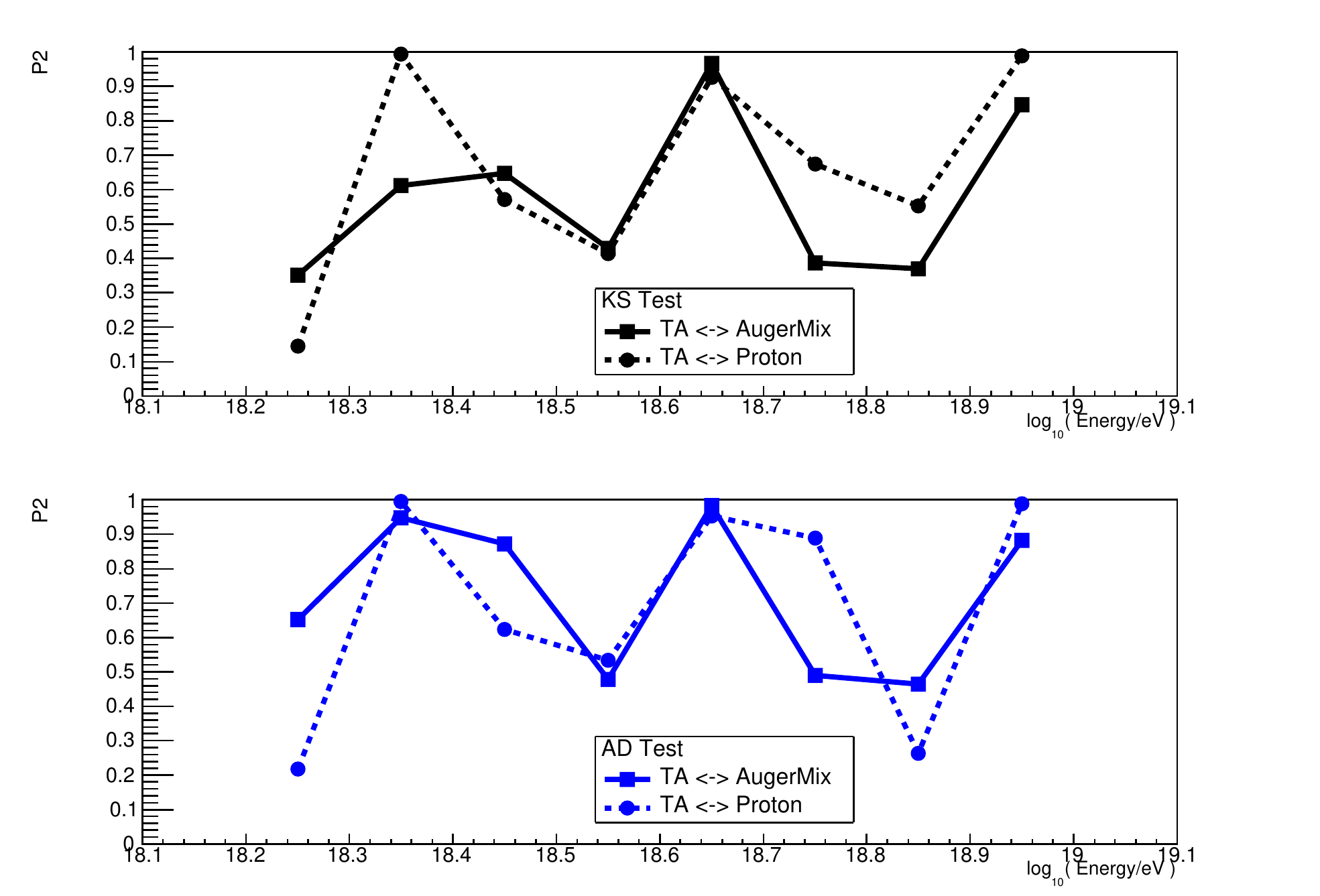}
  \caption{Compatibility probabilities (P2) between TA data and the
    Auger mix and QGSJet~II-04 protons, using the Kolmogorov-Smirnov
    test (top) and Anderson-Darling test (bottom). Probabilities $>
    0.01$ indicate acceptable agreement between TA data and the
    distribution under test.}
    \label{fig:auger_ta_compat}
\end{figure}

In addition, TA data was tested for compatibility with pure
QGSJet~II-04 protons and a similar level of agreement was
found. Probabilities greater than 0.01 were considered compatible. The
conclusion of the study was that, within systematic uncertainties,
TA's and Auger's data are compatible, and TA data is also compatible,
to about the same level, with a pure proton composition for $E <
10^{19}$~eV. Work is continuing with the Composition Working Group to
continue comparing data as both groups collect more
data. Figure~\ref{fig:auger_ta_compat} shows the computed
probabilities for all tests; table~\ref{tab:auger_ta_compat}
summarizes the results.

\begin{table}
  \small
  \centering
  \begin{tabular}{|rr|rrr|rrr|}
    \multicolumn{2}{c}{} &\multicolumn{3}{c}{Mix} &\multicolumn{3}{c}{Proton}\\
    \hline
    $E_{\mathrm{low}}$ &$E_{\mathrm{high}}$ &$\Delta$ &KS &AD
    &$\Delta$ &KS &AD\\
    \hline
    18.20 &18.30 &-23 &0.35 &0.65 &-31 &0.14 &0.21\\
    18.30 &18.40 &-26 &0.61 &0.95 &-33 &0.99 &0.99\\
    18.40 &18.50 &-16 &0.65 &0.87 &-22 &0.57 &0.62\\
    18.50 &18.60 &-12 &0.43 &0.48 &-21 &0.41 &0.53\\
    18.60 &18.70 &-12 &0.97 &0.98 &-24 &0.92 &0.95\\
    18.70 &18.80 &-6  &0.39 &0.49 &-20 &0.67 &0.88\\
    18.80 &18.90 &-15 &0.37 &0.47 &-31 &0.55 &0.26\\
    18.90 &19.00 &-4  &0.85 &0.88 &-20 &0.98 &0.98\\
    \hline
  \end{tabular}
  \caption{Compatibility probabilities (P2) between Auger and TA data,
    and QGSJet~II-04 protons and TA data using KS and AD two-sample
    tests. $\Delta$ is shift in \xm{} in g/cm$^{2}$ applied to the TA data.}
  \label{tab:auger_ta_compat}
\end{table}

\section{Summary}\label{sec:summary}
Telescope Array has recently published the first results of UHECR
\xm{} measurements for $E \geq 10^{18.2}$~eV, using the Black Rock and
Long Ridge FD stations in conjunction with the SD array. This nearly
nine year hybrid data set is TA's highest statistics measure of
\xm{}. \mxm{} and \sxm{} of the observed \xm{} distributions agree
with a model consisting of light composition. Above $10^{19}$~eV,
statistics of TA's hybrid reconstruction begin to fall to levels that
make testing data and Monte Carlo difficult. TA performed a test of
compatibility of observed \xm{} with four different primary chemical
elements using the QGSJet~II-04 hadronic model. For each energy bin
TA's observed \xm{} distributions were allowed to shift systematically
to find the shift which resulted in the largest maximum likelihood
calculated between data and Monte Carlo. The chance probability of
observing a likelihood at least as extreme as that found in the
shifted data was computed for all four elements. Using this procedure
it was found that TA data is compatible with QGSJet~II-04 protons for
all energies tested, $10^{18.2} \leq E < 10^{19.9}$~eV. Helium is
compatible for $E \geq 10^{19}$~eV, nitrogen is compatible for $E \geq
10^{19.2}$~eV, and iron is compatible for $E \geq 10^{19.4}$~eV.

Auger and Telescope Array continue to work together through a
Composition Working Group to allow the exchange of data and ideas to
resolve the apparent discrepancies in the interpretation of UHECR
composition. Auger and TA data cannot be directly compared because of
the different approaches to analysis each experiment takes. Auger
produces \xm{} distributions that are unbiased relative to true \xm{}
distributions produced in UHECR simulations. TA produces \xm{}
distributions which are biased due to detector acceptance and
resolution. Because of these differences in analysis procedures,
direct comparison of \mxm{} and \sxm{} does not provide an accurate
picture of what the interpretation of UHECR composition should
be. Instead, Auger and TA have developed a method of indirect
comparison of data. Auger produces an energy dependent ad-hoc mixture
of protons, helium, nitrogen, and iron that best fits their \xm{}
distributions. This mixture is then simulated by TA and reconstructed
in the normal manner. This causes the same biases to be applied to the
data observed by TA. Tests of the reconstructed mix can then be
performed to compare it to TA data. Work was done recently to test the
compatibility probability of Auger and TA data using this
procedure. It was found that TA data and Auger data are compatible
between $10^{18.2} \leq E < 10^{19.0}$~eV and the same level of
compatibility was also found, using the same tests, between TA data
and a pure proton composition. The Composition Working Group continues
to work together to improve their understanding of each group's
analysis and finding ways to compare their data.

\section*{Acknowledgements}
The Telescope Array experiment is supported by the Japan Society for
the Promotion of Science (JSPS) through 
Grants-in-Aid
for Priority Area
431,
for Specially Promoted Research 
JP21000002, 
for Scientific  Research (S) 
JP19104006, 
for Specially Promoted Research 
JP15H05693, 
for Scientific  Research (S)
JP15H05741 and
for Young Scientists (A)
JPH26707011; 
by the joint research program of the Institute for Cosmic Ray Research (ICRR), The University of Tokyo; 
by the U.S. National Science
Foundation awards PHY-0601915,
PHY-1404495, PHY-1404502, and PHY-1607727; 
by the National Research Foundation of Korea
(2016R1A2B4014967, 2016R1A5A1013277, 2017K1A4A3015188, 2017R1A2A1A05071429) ;
by the Russian Academy of
Sciences, RFBR grant 16-02-00962a (INR), IISN project No. 4.4502.13,
and Belgian Science Policy under IUAP VII/37 (ULB). The foundations of
Dr. Ezekiel R. and Edna Wattis Dumke, Willard L. Eccles, and George
S. and Dolores Dor\'e Eccles all helped with generous donations. The
State of Utah supported the project through its Economic Development
Board, and the University of Utah through the Office of the Vice
President for Research. The experimental site became available through
the cooperation of the Utah School and Institutional Trust Lands
Administration (SITLA), U.S. Bureau of Land Management (BLM), and the
U.S. Air Force. We appreciate the assistance of the State of Utah and
Fillmore offices of the BLM in crafting the Plan of Development for
the site.  Patrick Shea assisted the collaboration with valuable advice 
on a variety of topics. The people and the officials of Millard County, 
Utah have been a source of
steadfast and warm support for our work which we greatly appreciate. 
We are indebted to the Millard County Road Department for their efforts 
to maintain and clear the roads which get us to our sites. 
We gratefully acknowledge the contribution from the technical staffs of
our home institutions. An allocation of computer time from the Center
for High Performance Computing at the University of Utah is gratefully
acknowledged.

\bibliography{isvhecri2018_hanlon}

\end{document}